\documentstyle[12pt]{article}
\begin{document}
\pagestyle{empty}
\vspace* {13mm}
\renewcommand{\thefootnote}{\fnsymbol{footnote}}
\begin{center}
   {\bf A GENERALIZED DUALITY TRANSFORMATION OF THE \\[1mm]
	ANISOTROPIC $XY$ CHAIN IN A MAGNETIC FIELD} \\[25MM]
   Haye Hinrichsen\\[5mm]
   {\it Universit\"{a}t Bonn,
   Physikalisches Institut \\ Nu{\ss}allee 12,
   D-53115 Bonn, Germany}
\\[5cm]
{\bf Abstract}
\end{center}
\renewcommand{\thefootnote}{\arabic{footnote}}
\addtocounter{footnote}{-1}
\vspace*{2mm}
We consider the anisotropic $XY$ chain in a
magnetic field with special boundary conditions
described by a two-parameter Hamiltonian. It is shown that
the exchange of the parameters corresponds to a
similarity transformation, which reduces in a special
limit to the Ising duality transformation.
\vspace{4cm}
\begin{flushleft}
   BONN HE-93-40 \\
   cond-mat/9310055 \\
   October 1993 \\
\end{flushleft}
\thispagestyle{empty}
\mbox{}
%
%
%%%%%%%%%%%%%%%%%%%%%%%%%%%%%%%%%%%%%%%%%%%%%%%%%%%%%%%%%%%%%%%
%
%
\newpage
\setcounter{page}{1}
\pagestyle{plain}
\setcounter{equation}{0}
\indent
In this paper we consider the anisotropic $XY$ chain
in a magnetic field which is defined by the Hamiltonian
\begin{equation}
\label{TwoParameterHamiltonian}
\label{xy}
H^{XY}(\eta,q) \; = \;
-\frac{1}{2}\sum_{j=1}^{L-1}\:
\left( \,\eta\:\sigma_{j}^{x}\sigma_{j+1}^{x}
\; + \; \eta^{-1}\sigma_{j}^{y}\sigma_{j+1}^{y}
\; + \; q\,\sigma_{j}^{z} \; + \;
q^{-1}\sigma_{j+1}^{z} \right) \;,
\end{equation}
\noindent
where $q$ and $\eta$ are complex parameters
and $\sigma^{x,y,z}_j$
are Pauli matrices acting on site $j$.
Up to boundary terms, which will play a
crucial role here, $H$ can be rewritten as
\begin{equation}
\label{PeriodicHamiltonian}
H^{XY}(\eta,h) \; = \;
-\frac{1}{2}\sum_{j=1}^{L}\:
\left( \,\eta\:\sigma_{j}^{x}\sigma_{j+1}^{x}
\; + \; \eta^{-1}\sigma_{j}^{y}\sigma_{j+1}^{y}
\right)
\;-\; h\, \sum_{j=1}^{L}\; \sigma_j^z
\end{equation}
where $h=\frac{q+q^{-1}}{2}$ is the magnetic field.
This Hamiltonian
has a long history \cite{Nijs,McCoy} and provides
a good model for Helium adsorbed on metallic
surfaces ($\eta$ real and $q$ on the unit circle).
It also gives the master equation of the kinetic
Ising model \cite{Siggia} ($q=1$ and $\eta$ real) and
plays a role in one-dimensional reaction-diffusion
processes \cite{Chemie}. For the special boundary
conditions defined in Eq.~(\ref{xy}) the chain has
been shown to be invariant under a two-parameter
deformation of the $su(1|1)$ superalgebra \cite{Paper1},
and some of their correlation functions in the massless
regime have been computed in Ref. \cite{Paper2}.
\\
\indent
The aim of this paper is to show that for these boundary
conditions the exchange of the
parameters $q$ and $\eta$
in the Hamiltonian (\ref{xy})
corresponds to a similarity transformation
\begin{equation}
\label{SimH}
H^{XY}(\eta,q) \; \doteq \; H^{XY}(q,\eta)\;,
\end{equation}
which reduces in a special limit to the Ising duality transformation
(here `$\doteq$' denotes equality up to a similarity transformation).
The Ising limit of the $XY$ chain is given by
\begin{equation}
\label{limit}
H^{IS}(a,b) \;=\; \lim_{\xi \rightarrow \infty}
\,\frac{1}{\xi}\,H^{XY}(a\xi,b\xi)\,,
\end{equation}
\noindent
where
\begin{equation}
\label{HIsing}
H^{IS}(a,b) \;=\;
-\frac{1}{2}\sum_{j=1}^{L-1}\:
\left( \,a\:\sigma_{j}^{x}\sigma_{j+1}^{x}
\; + \; b\,\sigma_{j}^{z} \right) \;
\end{equation}
\noindent
is the Ising Hamiltonian with mixed boundary conditions \cite{mixed}.
One of the most remarkable properties of the Ising model is
its self-duality \cite{Kogut}. For the boundary conditions
defined in Eq. (\ref{HIsing}) the
Ising duality transformation
\begin{equation}
\label{DualTrans}
\sigma_j^x \rightarrow \tilde{\sigma}_j^x \;=\;
\prod_{i=1}^j\,\sigma_i^z\,,
\hspace{15mm}
\sigma^z_j \rightarrow  \tilde{\sigma}_j^z \;=\;
\sigma_j^x\,\sigma_{j+1}^x
\end{equation}
\noindent
takes place as
\begin{equation}
\label{IsDual}
H^{IS}(a,b) \;\doteq\; H^{IS}(b,a) \,+\, a\,(\sigma_L^z-\sigma_1^z)\,.
\end{equation}
\noindent
Using Eq. (\ref{limit}) we can rewrite Eq. (\ref{IsDual}) by
\begin{equation}
\label{eq6}
\lim_{\xi \rightarrow\infty} \frac{1}{\xi}\,H^{XY}(a\xi,b\xi)
\;\doteq\;
\lim_{\xi \rightarrow\infty} \frac{1}{\xi}\,H^{XY}(b\xi,
\frac{1}{a\xi})\,.
\end{equation}
\noindent
Notice that we absorbed the surface terms in Eq. (\ref{IsDual})
by inserting the argument $\frac{1}{a\xi}$ instead of~$a\xi$
on the r.h.s. of Eq. (\ref{eq6}). In order to symmetrize this
expression, we perform a rotation
\begin{equation}
\label{rotation}
\sigma_j^x \rightarrow \sigma^y_j\,,
\hspace{15mm}
\sigma_j^y \rightarrow -\sigma^x_j\,,
\hspace{15mm}
\sigma_j^z \rightarrow \sigma^z_j\,,
\hspace{20mm}
(j=1,\ldots,L)
\end{equation}
\noindent
on the l.h.s. of Eq. (\ref{eq6}):
\begin{equation}
H^{XY}(a\xi,b\xi) \;\doteq \; H^{XY}(\frac{1}{a\xi},b\xi)\,,
\end{equation}
\noindent
and we obtain
\begin{equation}
\label{eq9}
\lim_{\xi \rightarrow\infty} \frac{1}{\xi}\,H^{XY}
(\frac{1}{a\xi},b\xi)
\;\doteq\;
\lim_{\xi \rightarrow\infty} \frac{1}{\xi}\,H^{XY}(b\xi,
\frac{1}{a\xi})\,.
\end{equation}
\noindent
This means that Eq. (\ref{SimH}) holds for $\eta=\frac{1}{a\xi}$
and $q=b\xi$ in the limit $\xi \rightarrow \infty$.
\\[5mm]
\indent
The claim of this paper is to prove Eq. (\ref{SimH}),
i.e. we derive a similarity transformation
\begin{equation}
\label{udef}
H^{XY}(\eta,q) \;=\; U\,H^{XY}(q,\eta)\,U^{-1}
\end{equation}
\noindent
for {\it arbitrary} parameters $\eta$ and $q$.
Let us first summarize our results.
For that purpose let us introduce fermionic operators
by a Jordan-Wigner transformation
\begin{equation}
\label{jw}
\tau^{x,y}_j \;=\; \Bigl(\prod_{i=1}^{j-1}
\sigma_i^z\Bigr)\sigma_j^{x,y}
\end{equation}
\noindent
which allows the Hamiltonian (\ref{xy}) to be written as
\begin{equation}
\label{TauHamiltonian}
H(\eta,q)\;=\; \frac{i}{2}\sum_{j=1}^{L-1}
\left(
   \eta\,\tau_{j}^-\tau_{j+1}^+
   \ - \  \eta^{-1}\,\tau_{j}^+\tau_{j+1}^-
   \ + \ q\tau_{j}^+\tau_{j}^-
   \ + \ q^{-1}\tau_{j+1}^+\tau_{j+1}^- \,
\right).
\end{equation}
\noindent
Denoting
\begin{equation}
\label{abbr}
\alpha = \frac{q}{\eta} \,, \hspace{1.8cm}
\omega \;=\; \Bigl( \frac{\alpha^{1/2}-\alpha^{-1/2}}
{\alpha^{1/2}+\alpha^{-1/2}} \Bigr)
\end{equation}
\noindent
the explicit expression for $U(\alpha)$ is given by
the polynomial
\begin{equation}
\label{OrthogonalDefinition}
U(\alpha)\;=\;\frac1{\sqrt{N}}\,
\sum_{k=0}^{[L/2]}
\omega^k\, G_{2k}\,
\end{equation}
\noindent
where the generators $G_{2k}$ are defined by
\begin{equation}
\label{OrthoGenerators}
G_n \; = \sum_{1 \leq j_1<j_2<\ldots<j_{n} \leq L}
\tau^x_{j_1}\tau^x_{j_2}\ldots\tau^x_{j_{n}}\,.
\end{equation}
\noindent
By convention we take $G_0 \equiv {\bf 1}$, and
$[L/2]$ denotes the truncation of $L/2$ to an integer number.
$N$ is a normalization constant which is given by
\begin{equation}
\label{NormN}
N\;=\;\sum_{k=0}^{[L/2]}\,
{L \choose 2k}\,
\omega^{2k}
\;=\; 2^{L-1}\,\frac{1+\alpha^L}{(1+\alpha)^L}\;.
\end{equation}
Notice that the transformation depends only on the
ratio $\alpha=q/\eta$. Obviously
the normalization $N$ vanishes for $\alpha^L=-1$
so that the transformation (\ref{udef}) diverges,
and therefore
we will exclude this case in the following. For $\alpha=1$
the transformation $U(\alpha)$ reduces to the identity,
and this is what we expect
since for $\eta=q$ the exchange of $\eta$ and $q$ does not
effect a change in the Hamiltonian (\ref{xy}).
\\
\indent
In order to express the transformation in a formal way, let us
introduce the `time-ordered product'
\begin{equation}
T\,\tau_i^x\tau_j^x \;=\;
\left\{ \begin{array}{ll}
\hspace{2.3mm}\tau_i^x\tau_j^x\hspace{5mm}
& i<j \\
-\tau_j^x\tau_i^x & i>j \\
\hspace{5mm}0 & i=j
\end{array} \right\}\,,
\end{equation}
\noindent
which arranges the operators $\tau^x_j$ in increasing order with
respect to their fermionic commutation relations. Observing
that
\begin{equation}
G_{2k}=\frac{1}{k!}\,T\,G_2^k
\end{equation}
\noindent
where
\begin{equation}
G_2 \; = \sum_{1 \leq j_1<j_2 \leq L}
\tau^x_{j_1}\tau^x_{j_2}
\end{equation}
\noindent
we can rewrite
Eq. (\ref{OrthogonalDefinition}) formally as a time-ordered
exponential of $G_2$:
\begin{equation}
U(\alpha)\;=\;\frac{1}{\sqrt{N}}\,T\,\exp(\omega G_2)\,.
\end{equation}
\noindent
This expression suggests that the inverse of $U(\alpha)$ is
obtained by taking $\omega \rightarrow -\omega$, i.e.
$\alpha \rightarrow \alpha^{-1}$. In fact, one can show that
\begin{equation}
\label{Inverse}
U^{-1}(\alpha) \;=\; U(\alpha^{-1}) \,.
\end{equation}
\noindent
On the other hand we observe that $G_2^T = -G_2$ and thus
the transformation (\ref{udef}) is an orthogonal one:
\begin{equation}
\label{ortho}
U^T(\alpha) \;=\; U^{-1}(\alpha)\,.
\end{equation}
\noindent
It is interesting to know how the Pauli matrices
change under the transformation
\begin{equation}
\tilde{\sigma}_j^{x,y,z} \;=\; U(\alpha)\,
\sigma_j^{x,y,z}\, U^{-1}(\alpha)\,.
\end{equation}
\noindent
As we are going to show below, one obtains
three important identities:
\begin{eqnarray}
\label{ImportantId1}
\eta\,\tilde{\sigma}_j^x\tilde{\sigma}_{j+1}^x +
q\,\tilde{\sigma}^z_j
& = & q\,\sigma_j^x\sigma_{j+1}^x + \eta\,\sigma_j^z \\
\label{ImportantId2}
q\,\tilde{\sigma}_j^y\tilde{\sigma}_{j+1}^y +
\eta\,\tilde{\sigma}_{j+1}^z
& = & \eta\,\sigma_j^y\sigma_{j+1}^y + q\,\sigma_{j+1}^z \\
\tilde{\sigma}_j^x \tilde{\sigma}_{j+1}^y &=&
\sigma_j^x \sigma_{j+1}^y
\label{ImportantId3}\,.
\end{eqnarray}
\noindent
Because of these identities we have
\begin{eqnarray}
\label{SimilarityDefinition}
\label{QETransformation}
\tilde{H}^{XY}\,(q,\eta) \; = \; & -\frac{1}{2}\: \sum_{j=1}^{L-1}\:
[q\:\tilde{\sigma}_{j}^x\tilde{\sigma}_{j+1}^x
\; + \; q^{-1}\tilde{\sigma}_{j}^y\tilde{\sigma}_{j+1}^y
\; + \; \eta\,\tilde{\sigma}_{j}^z \; + \;
\eta^{-1}\tilde{\sigma}_{j+1}^z ]  & \\[3mm]
 \; = \; & -\frac{1}{2}\: \sum_{j=1}^{L-1}\:
[\eta\:\sigma_{j}^{x}\sigma_{j+1}^{x}
\; + \; \eta^{-1}\sigma_{j}^{y}\sigma_{j+1}^{y}
\; + \; q\,\sigma_{j}^{z} \; + \; q^{-1}\sigma_{j+1}^{z} ]
 & \;=\; H^{XY}(\eta,q)\,, \nonumber
\end{eqnarray}
\noindent
and our claim in Eq. (\ref{SimH}) is proved. The identities
(\ref{ImportantId1})-(\ref{ImportantId3}) contain even more
information. Since they hold independently
for every $1 \leq j < L$, it is obvious that
even the spectrum of the Hamiltonian
\begin{equation}
\label{ExtendedExchange}
\bar{H} \;=\;  -\frac{1}{2}\:  \sum_{j=1}^{L-1} \:
\left[\,a_j\,(\eta\:\sigma_{j}^{x}\sigma_{j+1}^{x}
\; + \; q\,\sigma_{j}^{z}) \;
\; + \; b_j\,(\eta^{-1}\:\sigma_{j}^{y}\sigma_{j+1}^{y}
\; + \; q^{-1}\:\sigma_{j+1}^{z})\,
\; + \; c_j\,\sigma_j^x\sigma_{j+1}^y \right]
\end{equation}
\noindent
is invariant under the exchange $q \leftrightarrow \eta$ for
{\it arbitrary} coefficients $a_j$, $b_j$ and $c_j$.
\\
\indent
We are now going to derive the identities
(\ref{ImportantId1})-(\ref{ImportantId3}).
For that purpose we first consider the transformation
properties of the fermionic operators
$\tilde{\tau}_j^{x,y}=U\,\tau_j^{x,y}\,U^{-1}$. It turns out
that
\begin{equation}
\label{UTransformation}
\tilde{\tau}_i^x \;=\; \sum_{i=1}^{L}\,
u_{i,j}\,\tau_j^x\,, \hspace{2cm}
\tilde{\tau}_j^y \;=\; \tau_j^y\,,
\end{equation}
\noindent
where
\begin{equation}
\label{UExpression}
u_{i,j}  \;=\;
\left\{
\begin{array}{ll}
\varrho \;\;\;\;\;\;\;\; & \mbox{if $i=j$}
\\[2mm]
\vspace{7mm}
(\varrho-\alpha)\,\alpha^{i-j} & \mbox{if $i<j$ }
 \\[-5mm]
(\varrho-\alpha^{-1})\, \alpha^{i-j} & \mbox{if $i>j$ }
\end{array}
\right\} \;,
\end{equation}
\noindent
and
\begin{equation}
\label{AbbrU}
\varrho\;=\;\frac{\alpha^{\frac{L}{2}-1}+\alpha^{-\frac{L}{2}+1}}
           {\alpha^{\frac{L}{2}}+\alpha^{-\frac{L}{2}}}\;.
\end{equation}
\noindent
Notice that the transformation (\ref{UTransformation})
is an orthogonal one as well
($\sum_{k=1}^L u_{i,k}u_{j,k} = \delta_{i,j}$) and reduces
for $\alpha=1$ to the identical transformation
$u_{i,j}=\delta_{i,j}$.
Furthermore the coefficients~$u_{i,j}$
depend only on the difference
of their indices $i-j$.
Obviously $\tau_j^y$ is invariant under the similarity
transformation, and this immediately
proves Eq. (\ref{ImportantId3}).
By adding the Jordan-Wigner
transformation (\ref{jw}) we obtain
the following transformation rules for other the terms
occuring in the Hamiltonian:
%
%
%%%%%%%%%%%%%%%%%%%%%%%%%%%%%%
%
%
\begin{eqnarray}
\label{Similarity1}
\tilde{\sigma}_j^x\tilde{\sigma}_{j+1}^x & =
& (\varrho-\alpha^{-1})\,\sum_{k=1}^{j-1}\alpha^{j-k+1}
    \sigma_k^y \, S_{k+1,j-1} \, \sigma_j^y \nonumber \\
&&+\,(\varrho-\alpha)\,\sum_{k=j+2}^L\alpha^{j-k+1}
    \sigma_j^x \, S_{j+1,k-1} \, \sigma_k^x \\[2mm]
&&+\,(1-\varrho\alpha)\,\sigma_{j}^z \;+\;
   \varrho\,\sigma_j^x\sigma_{j+1}^x \nonumber
\end{eqnarray}
\begin{eqnarray}
\label{Similarity2}
\tilde{\sigma}_j^y\tilde{\sigma}_{j+1}^y & =
& (\varrho-\alpha^{-1})\,\sum_{k=1}^{j-1}\alpha^{j-k}
    \sigma_k^y \, S_{k+1,j} \, \sigma_{j+1}^y \nonumber \\
&&+\,(\varrho-\alpha)\,\sum_{k=j+2}^L\alpha^{j-k}
    \sigma_{j+1}^x \, S_{j,k-1} \, \sigma_k^x \\[2mm]
&&+\,(1-\varrho\alpha^{-1})\,\sigma_{j+1}^z \;+\;
   \varrho\,\sigma_j^y\sigma_{j+1}^y \nonumber
\end{eqnarray}
\begin{eqnarray}
\label{Similarity3}
\tilde{\sigma}_j^z & =
&  \hspace{3mm}(\alpha^{-1}-\varrho)\,\sum_{k=1}^{j-1}\alpha^{j-k}
    \sigma_k^y \, S_{k+1,j-1} \, \sigma_{j}^y \\
&&+\,(\alpha-\varrho)\,\sum_{k=j+1}^L\alpha^{j-k}
    \sigma_{j}^x \, S_{j+1,k-1} \, \sigma_k^x \;\;
+\varrho\,\sigma_j^z , \nonumber
\end{eqnarray}
\begin{eqnarray}
\label{Similarity4}
\tilde{\sigma}_j^y\tilde{\sigma}_{j+1}^x & =
&  (\varrho-\alpha^{-1})\,\sum_{k=1}^{j-1}\alpha^{j-k}
   (\alpha\, \sigma_k^y \, S_{k+1,j} \, \sigma_{j+1}^x -
    \sigma_k^y \, S_{k+1,j-1} \, \sigma_{j}^x)\\
&&+\,(\varrho-\alpha)\,\sum_{k=j+2}^L\alpha^{j-k}
   (\sigma_j^y \, S_{j+1,k-1} \, \sigma_{k}^x - \alpha\,
    \sigma_{j+1}^y\,S_{j+2,k-1}\,\sigma_{k}^x)\\
&&+\,\Big(\varrho\,(\alpha+\alpha^{-1})-1\Bigr)\,
    \sigma_j^y\sigma_{j+1}^x , \nonumber
\end{eqnarray}
%
%
%%%%%%%%%%%%%%%%%%%%%%%%%%%%%%
%
%
\noindent
Here $S_{i,k}$ denotes the Jordan-Wigner string
between the sites $i$~and~$k$
\begin{equation}
\label{JWString}
S_{i,k} \;=\; \prod_{j=i}^{k}\,\sigma^z_j\;,
\hspace{2cm}
S_{i+1,i} \equiv 1 \;
\end{equation}
\noindent
which means that
the $q \leftrightarrow \eta$ transformation converts
local observables to linear combinations of
strings measuring the charge between certain positions.
Notice that Eqs. (\ref{Similarity1})-(\ref{Similarity4})
simplify in the thermodynamical limit $L \rightarrow \infty$,
where $\varrho \rightarrow \alpha$ if $|\alpha|<1$ and
$\varrho \rightarrow \alpha^{-1}$ if $|\alpha|>1$, respectively.
\\
\indent
We have discovered the identity (\ref{udef}) by first noticing
that the spectra of $H^{XY}(\eta,q)$
and $H^{XY}(q,\eta)$ are identical and we have computed the
relations (\ref{Similarity1})-(\ref{Similarity4}) by hand.
Then using
these results we conjectured the general structure of the
transformation (\ref{OrthogonalDefinition}).
\\
\indent
Let us finally check the Ising limit described above (c.f.
Eq.~(\ref{eq9})). For
$q \rightarrow \infty$ and $\eta \rightarrow 0$ Eqs.
(\ref{ImportantId1}) and~(\ref{ImportantId2}) reduce to
\begin{equation}
\label{NearlyDuality}
\tilde{\sigma}_i^z\;=\;\sigma_i^x\sigma_{i+1}^x \;,\;\;\;\;\;\;
\tilde{\sigma}_i^y\tilde{\sigma}_{i+1}^y\;=\;\sigma^z_{i+1}\;.
\end{equation}
\noindent
Now if we rotate $\tilde{\sigma}^x$ and  $\tilde{\sigma}^y$
like in Eq. (\ref{rotation}), we end up with
\begin{equation}
\label{FullDuality}
\tilde{\sigma}_i^z\;=\;\sigma_i^x\sigma_{i+1}^x \;,
\hspace{2cm}
\tilde{\sigma}_i^x\tilde{\sigma}_{i+1}^x\;=\;\sigma^z_{i+1}\;,
\end{equation}
\noindent
and this is just the Ising duality transformation
given in Eq. (\ref{DualTrans}).
\\
\indent
The $q \leftrightarrow \eta$ symmetry
(\ref{QETransformation}) may be interpreted physically as
follows: The parameter $q$ fixes (beyond the magnetic field)
the boundary conditions of the system while the parameter
$\eta$ describes the bulk anisotropy. So the exchange of $q$
and $\eta$ may be understood as a
transformation which exchanges the bulk and boundary
properties of the chain. An investigation of
correlation functions confirms this interpretation.\\[1cm]
{\bf Acknowledgements}\\
I would like to thank V. Rittenberg for valuable
discussions and
S. R. Dahmen for a careful reading of the manuscript.


\begin{thebibliography}{99}
\bibitem{Nijs}
    den Nijs N 1988
    {\it Phase Transitions and Critical Phenomena} \\
    ed Domb C and Lebowitz J L,
    vol. 12 (Academic Press New York) 219
\bibitem{McCoy}
    Barouch E, McCoy B M and Dresden M 1970
    {\it Phys. Rev.} {\bf A2} 1075 \\
    Barouch E and McCoy B M 1971
    {\it Phys. Rev.} {\bf A3} 786
\bibitem{Siggia}
    Siggia E D 1977 {\it Phys. Rev.} {\bf B16} 2319
\bibitem{Chemie}
    Alcaraz F C, Droz M, Henkel M and Rittenberg V 1992
    {\it Reaction-Diffusion Processes, Critical Dynamics and
    Quantum Chains} Geneva preprint UGVA-DPT 1992/12-799
\bibitem{Paper1}
   Saleur H 1989 {\it Trieste Conference
   on Recent Developements in
   Conformal Field Theories}
   ed Randjikar-Daemi, E. Sezgin and J. B. Zuber
   (World Scientific) 160\\
   Hinrichsen H and Rittenberg V 1992
   {\it Phys. Lett.} {\bf B 275} 350
\bibitem{Paper2}
   Hinrichsen H and Rittenberg V 1993
    {\it Phys. Lett.} {\bf B 304} 115
\bibitem{mixed}
   Cardy J L 1986 {\it Nucl. Phys.} {\bf B 275} 200
\bibitem{Kogut}
   Kogut J B 1979 {\it Rev. Mod. Phys} {\bf 51} 659
\end{thebibliography}
\end{document}